# Time varying ISI model for nonlinear interference noise

**Ronen Dar[1], Meir Feder[1], Antonio Mecozzi[2], and Mark Shtaif[1]**

[1]*School of Electrical Engineering, Tel Aviv University, Israel*
[2]*Department of Phyical and Chemical Sciences, University of L'Aquila , Italy*
*Tel Aviv 69978, ISRAEL*
*shtaif@eng.tau.ac.il*

**Abstract:** We show that the effect of nonlinear interference in WDM systems is equivalent to slowly varying inter-symbol-interference (ISI), and hence its cancelation can be carried out by means of adaptive linear filtering. We characterize the ISI coefficients and discuss the potential gain following from their cancellation.

**Introduction**

The capacity of fiber-optic communications systems is bounded due to the presence of nonlinear interference effects [1]. Since in a complex network setting different WDM channels are often routed through different and unpredictable paths, it is customary to treat nonlinear interference between them as noise. This noise (to which we refer as nonlinear interference noise – NLIN) is very frequently treated as additive Gaussian [2–4] and system performance is extracted solely from its variance. As was pointed out in [5,6], the nonlinear interference noise is not additive in reality, but rather it contains significant information on the channel of interest. Extraction of this information has the potential of significantly enhancing the performance of the fiber-optic channel. In a recent paper [7] we have shown that by taking advantage of the temporal correlations in the phase-noise component of the NLIN, the spectral efficiency can be increased by nearly 1 bit/sec/Hz relative to the predictions made in [1]. Here we generalize this concept and show that a channel with NLIN is rigorously equivalent to a linear channel with time-varying inter-symbol interference (ISI). In other words, the $n$-th received sample in the channel of interest can be written as $r_n = a_n + \Delta a_n$, where $a_n$ is the data symbol transmitted over the channel of interest, and where

$$\Delta a_n = \sum_k h_k^{(n)} a_{n-k} \qquad (1)$$

is the NLIN. The ISI coefficients $h_k^{(n)}$ depend on the data transmitted in the neighboring (interfering) WDM channels in a way that will be specified in what follows. Their dependence on time (which is reflected by the presence of the superscript $n$ in $h_k^{(n)}$) will be shown to be very slow on the scale of a symbol duration and hence they can be estimated and tracked based on real-time measurements of the channel of interest. Knowledge of the coefficients $h_k^{(n)}$ implies that rather than treating the error signal $\Delta a_n$ as noise, one can extract the information on the data symbols from $\Delta a_n$ by applying standard signal processing methods (such as sequential decoding, or adaptive decision feedback equalization), and significantly reduce the detection error. We note that the zero-th coefficient, $h_0^{(n)}$, has the form $h_0^{(n)} = i\theta_n$, where $\theta_n$ is the phase-noise coefficient which was presented in [5],[6] and whose cancellation was the subject of the study in [7],[8].

We provide simple closed form expressions for the mean-square values of the coefficients $h_k^{(n)}$, which allows us to evaluate the overall NLIN power

$$\langle |\Delta a_n|^2 \rangle = \langle |a_n|^2 \rangle \sum_k \langle |h_k^{(n)}|^2 \rangle, \qquad (2)$$

as well as the NLIN power that remains after some of the ISI contributions are eliminated by means of adaptive equalization. For example, upon cancellation of phase-noise [7], which is represented by the zeroth coefficient $h_0^{(n)}$, the remaining NLIN power is obtained by summing Eq. (2) over all indices $|k| \geq 1$. This result completes the analyses reported in [5],[6] by providing explicit analytical expressions not only for the phase-noise itself, but also for the noise that remains after its cancellation. We show in what follows that the contributions of the various terms in Eq. (2) to the NLIN reduce as $k^{-2}$ and that compensation for only three coefficients ($k = -1,0,1$), leads to the reduction of the nonlinear noise variance by more than an order of magnitude, whereas cancellation of only the zeroth term (phase-noise) reduces the NLIN power by a factor of 3.5 [6].

The analytical part of our work focuses on NLIN due to cross-phase modulation (XPM), which is known to be the dominant inter-channel nonlinearity in many systems of practical interest. The excellent agreement between our theory and simulations (which include all nonlinear effects) supports this assumption. Additionally, since in this work we are concerned with introducing the principle of nonlinearity cancelation, the analytical and numerical results that we report correspond to a system with perfectly distributed gain. This is the same scenario as the one in which the capacity limits of the nonlinear fiber-optic channel have been assessed [1], and in this case the performance depends only on the overall length of the system independently of the number of spans that it is divided into. However, we stress that the principle reflected by Eq. (1) holds in general, regardless of the assumed amplification profile.

**Theory and results**

The detection error due to XPM effects was analyzed in [5] where (for the case of a single interfering channel) it was shown to have the form

$$\Delta a_n = 2i\gamma \sum_{k,l,m} a_{n-k} b_{n-l}^* b_{n-m} X_{k,l,m}, \tag{3}$$

where $\gamma$ is the nonlinearity coefficient, $\{b_j\}$ are the data symbols in the interfering channel and $X_{k,l,m}$ is a coefficient [5],[6] whose value is defined by the waveforms of the individual pulses, the channel separation, the dispersion coefficient, and the length of the optical link. In the case of multiple interfering channels, $\Delta a_n$ is equal to the sum of the individual contributions. Equation (1) follows immediately by rearranging the coefficients and identifying $h_k^{(n)} = 2i\gamma \sum_{l,m} b_{n-l}^* b_{n-m} X_{k,l,m}$. We note that (as can be seen from [5]) the zeroth-order coefficient $h_0^{(n)} = i\theta_n$ with $\theta_n$ being phase-noise. Our goal is to characterize the statistics of the ISI coefficients $h_k^{(n)}$ (mean, variance and autocorrelation), a procedure that necessitates the evaluation of $X_{k,l,m}$, which is very challenging computationally since a large number of these coefficients needs to be found. A good approximation can be obtained in the case of Nyquist pulses with a square power spectrum of width $B$. While the analytical details of this approximation exceed the scope of this paper and will be provided separately, it allows the assessment of the means and variances of the coefficients $h_k^{(n)}$. For the zeroth coefficient $h_0^{(n)} = i\theta_n$, the mean and variance have been given in [5,6]. In the case of higher order ISI coefficients ($h_k^{(n)}$ for $|k| \geq 1$), the mean can be shown to be zero and the variance is given by

$$\langle |h_k|^2 \rangle = \left(\frac{2\gamma^2 L}{|\beta''|\pi^3}\right) \frac{1}{k^2} \sum_s [A(x_s)(\langle |b_n|^4 \rangle - \langle |b_n|^2 \rangle^2) + (A(x_s) + B_k(x_s))\langle |b_n|^2 \rangle^2], \tag{4}$$

where the superscript ($n$) was omitted since stationarity implies that the variance $\langle |h_k^{(n)}|^2 \rangle$ depends only on the value of $k$. The coefficients appearing in the equation are $A(x) = 1 - \frac{x+1}{2}\ln\left(\frac{x+1}{x}\right) + \frac{x-1}{2}\ln\left(\frac{x-1}{x}\right) \simeq \frac{1}{6x^2}$ and $B_k(x) = \left(\frac{1}{x^2} + \frac{1}{x^4} + \frac{1}{x^6} + \frac{1}{x^8}\right)\left(\frac{1}{6} - \frac{1}{\pi^2 k^2}\right) \simeq \frac{1}{6x^2}$. The summation in Eq. (4) is over all neighboring channels, and $x_s = B/|\Omega_s|$ with $\Omega_s$ being the frequency separation between the channel of interest and the $s$-th neighboring channel. Substitution of $\langle |h_k|^2 \rangle$ into Eq. (2) and summation over all $|k| \geq 1$ yields the NLIN that remains after nonlinear phase-noise is eliminated by means of an efficient phase-recovery algorithm. Summation over $|k| \geq k_0$ gives the NLIN variance that remains after both the phase-noise as well as ISI from $k_0$ nearest symbols has been eliminated by a proper algorithm. It is important to note that only the first phase-noise component of NLIN (the term that is proportional to $\langle |b_n|^4 \rangle - \langle |b_n|^2 \rangle^2$) is proportional to $\Omega_s^{-1}$, which implies logarithmic growth with the number of channels. All other terms depend on $\Omega_s^{-2}$, implying that the dependence on the number of WDM channels saturates rapidly.

**Numerical validation**

Validation of the theoretical results was performed in a set of split-step Fourier simulations. The simulations were performed for a 500 km system over a standard single mode fiber in a setting that is similar to the one used in [1] for the estimation of channel capacity. Specifically, we assumed a dispersion coefficient of 21 ps$^2$/km, a nonlinearity coefficient $\gamma = 1.3$ W$^{-1}$km$^{-1}$, a baud-rate of 100 Gbaud/s, and a channel spacing of 102 GHz. Nyquist pulses of a perfectly square optical spectrum (of 100 GHz width) were used. The number of simulated symbols in each run was 8192 and up to 500 runs (each with independent and random data symbols) were performed with each set of system parameters, so as to accumulate sufficient statistics. The data symbols of the various channels were generated independently. Use of very long sequences in every run is critical in such simulations so as to achieve acceptable accuracy in view of the long correlation time of NLIN, as well as to avoid artifacts related to the periodicity of the signals that is imposed by the use of the discrete Fourier transform. We simulated up to 5 WDM channels (as specified in the figures). At the receiver the channel of interest was isolated with a matched optical filter and back-propagated so as to eliminate intra-channel effects and chromatic dispersion.

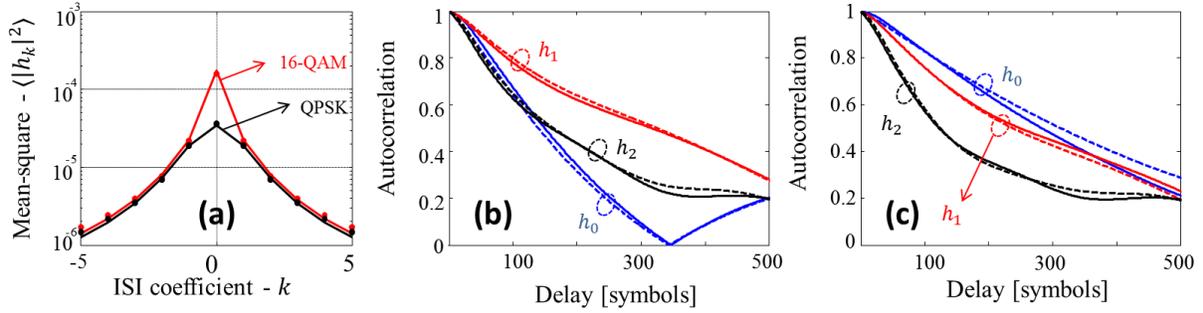

**Fig. 1. (a)** Mean-square values of ISI coefficients $\langle |h_k|^2 \rangle$ for QPSK and 16-QAM transmission. Solid lines show the theory (Eq. 4), dots show simulation results. **(b)** The autocorrelation functions of $h_0, h_1$, and $h_2$ for QPSK modulation. The dashed curves show analytical results (whose form will be detailed elsewhere), whereas the solid curves were obtained from the simulations. **(c)** same as (b) but for 16-QAM.

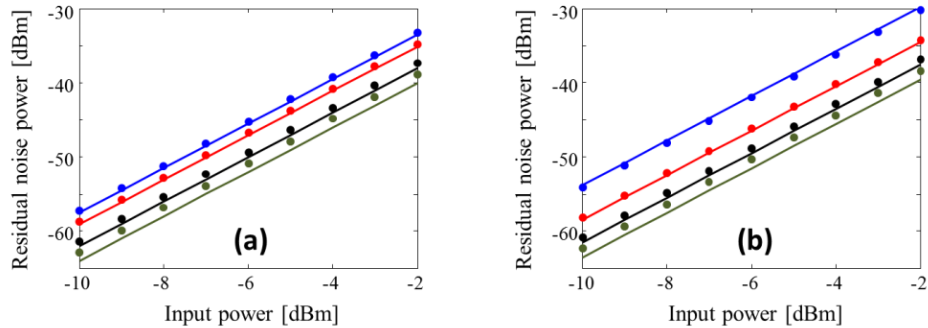

**Fig. 2. (a)** The NLIN power as a function of the input power for QPSK modulation without noise cancellation (blue curve), after cancellation of phase-noise ($h_0$, red), after cancellation of three ($h_0, h_{\pm 1}$, black), and five ($h_0, h_{\pm 1}, h_{\pm 2}$ green) ISI terms. Solid lines represent the theory and dots represent simulation results. **(b)** same as (a) but for 16-QAM.

In Figure 1 we validate the analysis by comparing the mean square value of the ISI coefficients $\langle |h_k|^2 \rangle$ given by Eq. (5) with the values extracted from the simulations. The excellent agreement between theory and simulation is self-evident. The extraction of $h_k^{(n)}$ from the simulations is based on the least-squares estimation method with a window size of 50 symbols (similar to [7]). The ability to extract the coefficients numerically relies on their long temporal correlation, which can be seen in Figs. 1b and 1c, where the autocorrelation functions of $h_0, h_1$, and $h_2$ are plotted. The dashed and solid curves correspond to the theoretical results and the results obtained from simulations, respectively. We note that for each $k$ the coefficients $h_k$, and $h_{-k}$ share the same autocorrelation function.

The potential effectiveness of nonlinear ISI cancellation is illustrated in Fig. 3. The curves and the symbols represent the results of the analysis and simulations, respectively. The blue curve shows the total noise power, without ISI cancellation. The red curve shows the remaining noise power after the contribution of the zeroth ISI coefficient $h_0$ (i.e. the phase-noise) is compensated for. The black curve shows the noise after cancellation of $h_0$ and $h_{\pm 1}$, and the green curve corresponds to the case where the effects of $h_0, h_{\pm 1}$ and $h_{\pm 2}$ are compensated for. In the case of QPSK modulation the potential gain from compensating for all five coefficients is 5.6dB, whereas the gain compensating only for $h_0$ (the phase-noise) is 1.6dB. In the case of 16-QAM the gains are 8dB and 4dB for compensation of all five coefficients and only $h_0$, respectively (slightly less than obtained with Gaussian modulation [7]).

**Conclusions**

We have shown that nonlinear interference between WDM channels manifests itself as a time varying ISI. The ISI coefficients have been expressed and characterized analytically. Their slow dependence on time allows their extraction from the received data, making compensation for the nonlinear interference possible. Potential gains of 5.6dB and 8dB have been demonstrated in the cases of QPSK and 16-QAM modulation.